\documentclass[pra,aps,twocolumn,nopacs,superscriptaddress]{revtex4}
\usepackage{amsmath,graphicx,epsfig}

\begin{document}

\title{Liquid-state nuclear spin comagnetometers} 

\author{M. P. Ledbetter}\email{micah.ledbetter@gmail.com}
\affiliation{Department of Physics, University of California at
Berkeley, Berkeley, California 94720-7300, USA}
\author{S. Pustelny}
\affiliation{Department of Physics, University of California at
Berkeley, Berkeley, California 94720-7300, USA}
\affiliation{Center for Magneto-Optical Research, Institute of Physics, Jagiellonian University, Reymonta 4, PL-30-059 Krak\'ow, Poland}
\author{D. Budker}
\affiliation{Department of Physics, University of California at
Berkeley, Berkeley, California 94720-7300, USA}
\affiliation{Nuclear Science Division, Lawrence Berkeley National
Laboratory, Berkeley CA 94720}
\author{M. V. Romalis}
\affiliation{Department of Physics, Princeton University, Princeton,
New Jersey 08544, USA}
\author{J. W. Blanchard}
\affiliation{Department of Chemistry, University of California at
Berkeley, Berkeley, California 94720-3220, USA}
\author{A. Pines}
\affiliation{Department of Chemistry, University of California at
Berkeley, Berkeley, California 94720-3220, USA}
\affiliation{Materials Science Division, Lawrence Berkeley National
Laboratory, Berkeley CA 94720}

\date{\today}


\begin{abstract}
We discuss nuclear spin comagnetometers based on ultra-low-field nuclear magnetic resonance in mixtures of miscible solvents, each rich in a different nuclear spin.  In one version thereof, Larmor precession of protons and ${\rm ^{19}F}$ nuclei in a mixture of thermally polarized pentane and hexafluorobenzene is monitored via a sensitive alkali-vapor magnetometer.  We realize transverse relaxation times in excess of 20 s and suppression of magnetic field fluctuations by a factor of 3400.  We estimate it should be possible to achieve single-shot sensitivity of about $5\times{\rm 10^{-9}~Hz}$, or about $5\times 10^{-11}~{\rm Hz}$ in $\approx 1$ day of integration. In a second version, spin precession of protons and ${\rm ^{129}Xe}$ nuclei in a mixture of pentane and hyperpolarized liquid xenon is monitored using superconducting quantum interference devices. Application to spin-gravity experiments, electric dipole moment experiments, and sensitive gyroscopes is discussed.
\end{abstract}

\maketitle

Atomic comagnetometers based on overlapping ensembles of different spins form the basis for many high-precision tests of fundamental symmetries \cite{Venema1992,Bear2000,Rosenberry2001,Regan2002,Griffith2009,Gemmel2010} and sensitive gyroscopes \cite{Kornack2005}.  Most such devices employ spins in gas-phase systems.  Here we demonstrate a new class of comagnetometers based on overlapping ensembles of nuclear spins in liquid state. The technique has the potential to strengthen limits on spin-gravity coupling and  electric dipole moments by several orders of magnitude.  Application to inertial sensing is also discussed.

The comagnetometer described here is similar in principle to proton-precession magnetometers \cite{Packard1954,Stuart1972,Waters1958}, in which nuclear spins, thermally polarized in a pulsed electromagnet, are allowed to precess in the field of interest.  Precession is detected via inductive pickup coils, and the precession frequency serves as a direct measure of the magnetic field. Our technique is based on ultra-low-field nuclear magnetic resonance of binary mixtures of mutually miscible solvents, each rich in a different nuclear spin-species. Rather than using inductive pickup coils, as in traditional proton precession magnetometers, we use sensitive alkali-vapor or superconducting quantum interference device (SQUID) magnetometers to probe nuclear spin-precession, enabling measurements to be performed with high signal-to-noise ratio at low magnetic fields $(\approx 1~{\rm mG})$ suitable for precision measurements.  Our discussion is primarily focused on a proton-${\rm ^{19}F}$ comagnetometer in a mixture of pentane and hexafluorobenzene (HFB), in which spins are thermally prepolarized in a strong magnetic field.  Spin precession is then monitored via an alkali-vapor magnetometer in a $\approx 1~{\rm mG}$ field.  We realize $T_2^\star$ of 13.7 and 20.8 s for ${\rm ^1H}$ and ${\rm ^{19}F}$, respectively.  Based on signal-to-noise projections for realistic conditions, we estimate that frequency resolution of about $5\times 10^{-9}~{\rm Hz}$ can be achieved in a single shot, or about $10^{-11}~{\rm Hz}$ in approximately one day of integration.  Sensitivity limits imposed by spin-projection noise are several orders of magnitude smaller.

We also briefly mention operation of a proton-${\rm ^{129}Xe}$ comagnetometer using a mixture of hyperpolarized liquid xenon and pentane.  In this version, SQUIDs are used to monitor nuclear spin precession in 10 mG magnetic fields.  In the Xe-pentane comagnetometer, $T_2$ is 3.5 s for protons and $\approx 250~{\rm s}$ for ${\rm ^1H}$ and ${\rm ^{129}Xe}$, respectively.

\begin{figure}
  \includegraphics[trim = 0in 0in 0in 3.5in, clip, width = 3.4in]{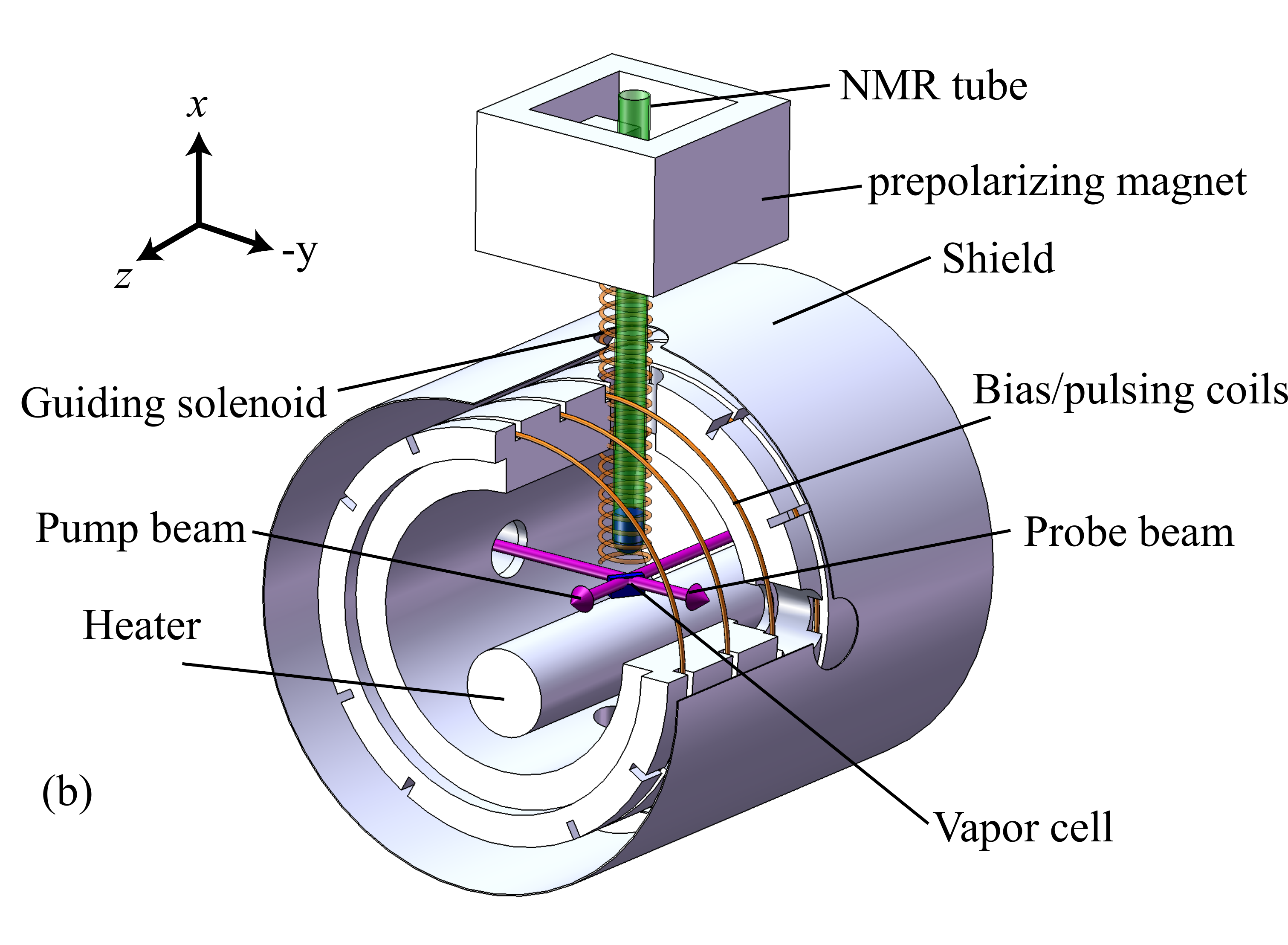}\\
  \caption{Experimental setup for the pentane-HFB nuclear-spin comagnetometer.  A sample of pentane and hexafluorobenzene is thermally polarized in a 20 kG magnet and moved into a low-field detection region in the presence of a ``guiding" field along the $x$ direction.  After the guiding field is removed, the spins precess in a $z$ directed magnetic field of $\approx 1~{\rm mG}$.  This precession is monitored by an alkali-vapor magnetometer operating in the SERF regime. }\label{Fig:ExpSetup}
\end{figure}

The experimental setup used for the pentane-HFB comagnetometer is similar to that of Ref. \cite{Ledbetter2011} and is shown in Fig. \ref{Fig:ExpSetup}.  Solvent mixtures (roughly 100 ${\rm \mu L}$) were degassed via five freeze-thaw cycles under vacuum and flame sealed in a 5 mm NMR tube.  Samples were thermally polarized in a 20~kG magnet and pneumatically shuttled into a low-field ($B_z\approx 1~{\rm mG}$) detection region adjacent to an alkali-vapor atomic magnetometer.  A set of magnetic shields isolates the system from ambient magnetic field fluctuations, and a set of coils provides control over the magnetic fields and gradients. A solenoid generates a ``guiding" field during the sample transit, so that the initial magnetization is along the $x$ direction. After the sample has arrived in the low-field region, the guiding field is removed and the nuclear spins precess about $B_z$.

The optical-atomic magnetometer consists of a microfabricated alkali-vapor cell; $z$-directed, circularly polarized pump light tuned to the center of the D1 transition; and $y$-directed, linearly polarized probe light tuned about 100 GHz to the blue of the D1 transition.  In the relatively small magnetic fields used in this work, relaxation of the alkali polarization, due to spin-exchange collisions is mostly eliminated \cite{Happer1973,Kominis2003}.  In this way, magnetometric sensitivity of about ${\rm 1.6~pG/Hz^{1/2}}$ has been achieved \cite{Kornack2007}.  Here we use much smaller vapor cells, and application of ${\rm \approx mG}$ magnetic fields broadens the alkali magnetometer resonance, resulting in low-frequency sensitivity of about ${\rm 2~nG/Hz^{1/2}}$.  Incorporation of a solenoid around the nuclear spin sample would enable independent control of the magnetic field applied to the sample and alkali-vapor \cite{Yashchuk2004}, reducing such broadening.  More details of the magnetometer operation can be found in the Supplementary Information.

Figure \ref{Fig:pentane_HFB} shows the single-shot ultra-low-field NMR signal obtained in a mixture of pentane and hexafluorobenzene (roughly equal volumes). The signal is well described by a sum of two exponentially decaying sinusoids, with transverse relaxation time $T_2^\star = 13.7$ and 20.8 s for protons and ${\rm ^{19}F}$ nuclei.  Obtaining such long transverse relaxation times required careful compensation of magnetic field gradients.  Uncertainties in the frequencies extracted from the fit are typically on the order of 30-70 ${\rm \mu Hz}$.    The inset in Fig. \ref{Fig:pentane_HFB} shows the real part of the Fourier transform.  The phase of the signal is a result of sensitivity to magnetic field in both the $x$ and $y$ direction in the presence of $B_0$, and is discussed in the Supplementary Information.  Similar signals were also obtained in a mixture of hexafluorobenzene and acetone (see Supplementary Information).

It is interesting to note that neat cyclopentane had $T_2^\star=11~{\rm s}$ (probably limited by gradients) and neat benzene had $T_2^\star=21~{\rm s}$ (after optimizing gradients), however, mixtures of hexafluorobenzene and cyclopentane or benzene yielded fast relaxation, with $T_2\approx 0.8~{\rm s}$. This may be due to strong intermolecular interactions between benzene and hexafluorobenzene \cite{Patrick1960}, or possibly due to residual oxygen.  For reference, signals from neat pentane, cyclopentane, benzene, tetramethylsilane and ${\rm ^{13}C}$ labeled formic acid, are presented in the Supplementary Information.

\begin{figure}
  \includegraphics[width=3.4in]{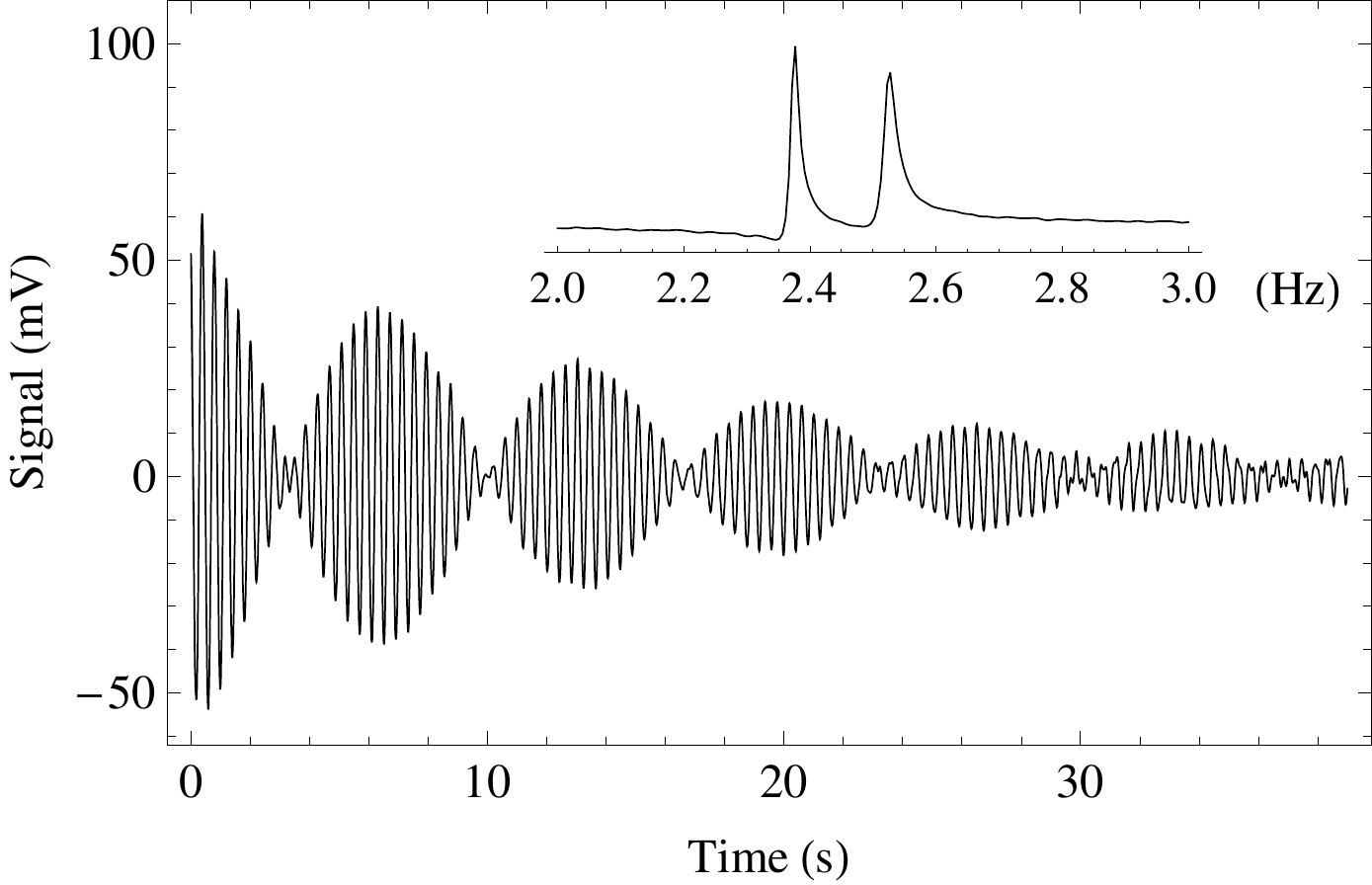}\\
  \caption{Ultra-low-field ($0.6~{\rm mG}$) NMR signal (single shot) from a mixture of pentane and hexafluorobenzene. Data are well described by the sum of two exponentially decaying sinusoids with $T_2 = 13.7$ and 20.8 s. The inset shows the real part of the Fourier transform.  }\label{Fig:pentane_HFB}
\end{figure}

\begin{figure}
  \includegraphics[trim = 0.7in 0in 0.7in 0in, clip, width=3.4in]{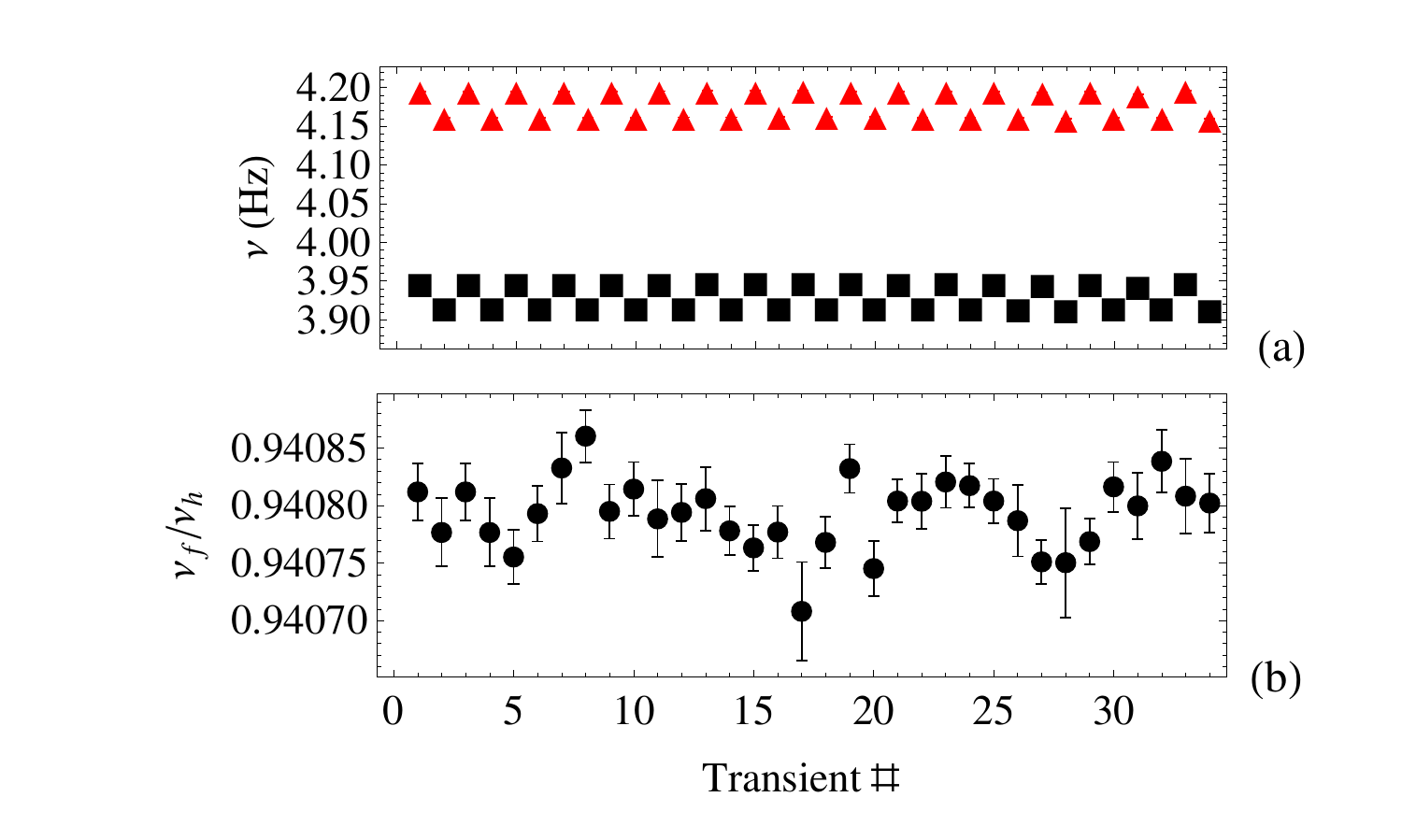}\\
  \caption{Demonstration of operation as a comagnetometer. (a) Free precession frequency of ${\rm ^1H}$ (triangles) and ${\rm ^{19}F}$ (squares) nuclei is shown as a small modulation is applied to $B_z$, $\Delta B_z\approx 0.002~{\rm mG}$, demonstrating that the two frequencies track each other closely.  (b) Modulation is not visible in the ratio $\nu_f/\nu_h$.}\label{Fig:field_tracking}
\end{figure}

As a demonstration of operation as a comagnetometer, we present in Fig \ref{Fig:field_tracking}(a) the frequency of ${\rm ^1H}$ and ${\rm ^{19}F}$ nuclei in a mixture of pentane and hexafluorobenzene for 32 transients. Every other transient, the magnetic field switches between $B_0+0.0037~{\rm mG}$ and $B_0-0.0037~{\rm mG}$ ($B_0 = 950~{\rm \mu G}$) corresponding to a peak-to-peak modulation of about 1 part in 120.  The two frequencies track each other well. To characterize the degree to which magnetic field fluctuations may be compensated, we plot the ratio $\nu_f/\nu_h$ in Fig. \ref{Fig:field_tracking}(b), where there is no apparent modulation. Peak-to-peak amplitude of $\nu_f/\nu_h$ at the magnetic field modulation frequency (determined using a software lock-in) is at the level of 1 part in 420000, representing suppression of magnetic field noise by a factor of about 3400.

\begin{figure}
  \includegraphics[trim = 0in 0in 0in 0in, clip, width=3.4in]{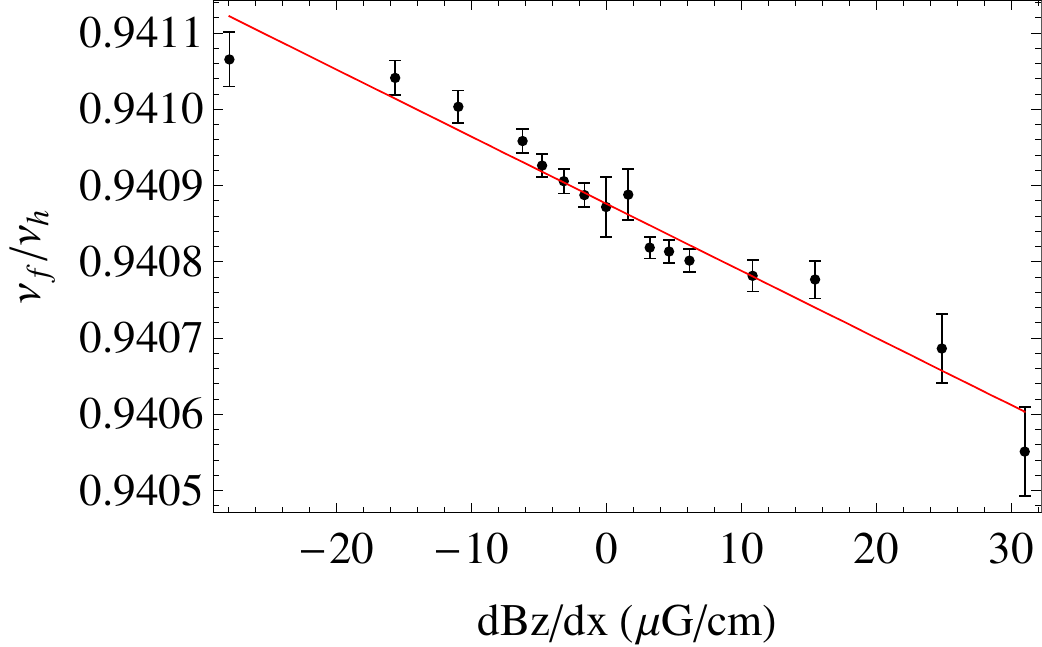}\\
  \caption{Effects of gradients on the pentane-HFB comagnetometer.  A gradient, $dB_z/dx$ was applied and the ratio $\nu_f/\nu_h$ is plotted.  The variation in the ratio of the frequencies indicates that there is some separation of the two solvents .}\label{Fig:gravity_gradients}
\end{figure}

The data in Fig. \ref{Fig:field_tracking}(b) are consistent with literature values for the magnetic moments of protons and ${\rm ^{19}F}$ \cite{CRC}, which give $\nu_f/\nu_h = 0.94077$, however they display long term drift in the ratio that is somewhat larger than the errors in individual measurements.  A likely explanation for this is spin-density gradients due to the different densities of the solvents (0.62 g/cc for pentane and 1.62 g/cc for HFB) in conjunction with drifting magnetic field gradients or jitter in the position of the sample in the presence of a static gradient.  A small degree of separation between the ``center of spin" of the two spin species is experimentally confirmed by examining the dependence of $\nu_f/\nu_h$ on the gradient $g_x=dB_z/dx$ (Fig. \ref{Fig:gravity_gradients}).  $\nu_f/\nu_h$ is roughly linear in the gradient, with a slope $-8.8\times 10^{-6}~{\rm cm/\mu G}$.  If the centers of spin for the two ensembles is separated by an amount $\Delta$, to first order in $g_x \Delta$, the ratio $\nu_f/\nu_h = (\gamma_f/\gamma_h)(1-g_x\Delta/B_0)$. From the observed dependence on gradients, we establish that $\Delta\approx 0.0085~{\rm cm}$. The RMS drift in the ratio $\nu_f/\nu_h$ shown in Fig. \ref{Fig:field_tracking}(b) is about 0.00003.  This would be accounted for by gradient drift of about $3~{\rm \mu G/cm}$. This seems large, though not inconceivable, given that the sample is being shuttled up and down. Sensitivity to gradients could be reduced by establishing motional narrowing, either via convection, or by mechanically spinning the sample.  A more detailed discussion of systematic effects will be presented elsewhere.

The ultimate sensitivity of precision measurements based on a thermally polarized pentane-HFB sample can be estimated as follows: Assuming sensor noise described by $\rho$, the accuracy with which one can determine the precession frequency in a single measurement of duration $T_2$, is roughly equal to $\delta\nu = \rho/(2\pi B_s T_2^{3/2})$. Assuming $r = 2r_0$, the signal amplitude is $B_s=\pi M_j/3$, where $M_j = n_j\mu_j^2B_p/kT$.  For equal-volume mixtures of pentane and HFB, the ${\rm ^1H}$ and ${\rm^{19}F}$ densities are $n_h = 3.1\times 10^{22}~{\rm cm^{-3}}$ and $n_f = 1.6\times 10^{22}~{\rm cm^{-3}}$, and a polarizing field of 100 kG at room temperature, $B_s = 16$ and 7~${\rm \mu G}$ for ${\rm ^1H}$ and ${\rm ^{19}F}$, respectively. (We verified that signal amplitudes in the present experiment are roughly consistent with those expected from thermal polarization, see Supplementary Information.) SQUIDs and atomic magnetometers are both capable of reaching sensitivities below $10~{\rm pG/Hz^{1/2}}$.  Using these numbers, and $T_2=10~{\rm s}$, we find that the single shot frequency resolution is about $\delta\nu=3\times 10^{-9}$ and $7\times 10^{-9}~{\rm Hz}$ for ${\rm ^1H}$ and ${\rm ^{19}F}$, respectively.  Averaging $10^4$ such measurements for a total integration time of roughly one day improves these numbers by a factor of 100.  This estimate neglects spin-projection noise as it is several orders of magnitude smaller than that imposed by the finite sensitivity of the magnetometers.

In the context of magnetic-field measurements, electron spins are often more appropriate because of the larger gyromagnetic ratio.  Nevertheless, frequency resolution of $3\times 10^{-9}~{\rm Hz}$ for protons corresponds to magnetic-field resolution at the level of 0.5 pG, approaching the sensitivity of the state-of-the-art SQUID and alkali-vapor atomic magnetometers.

\begin{figure}
  \includegraphics[width=3.4 in]{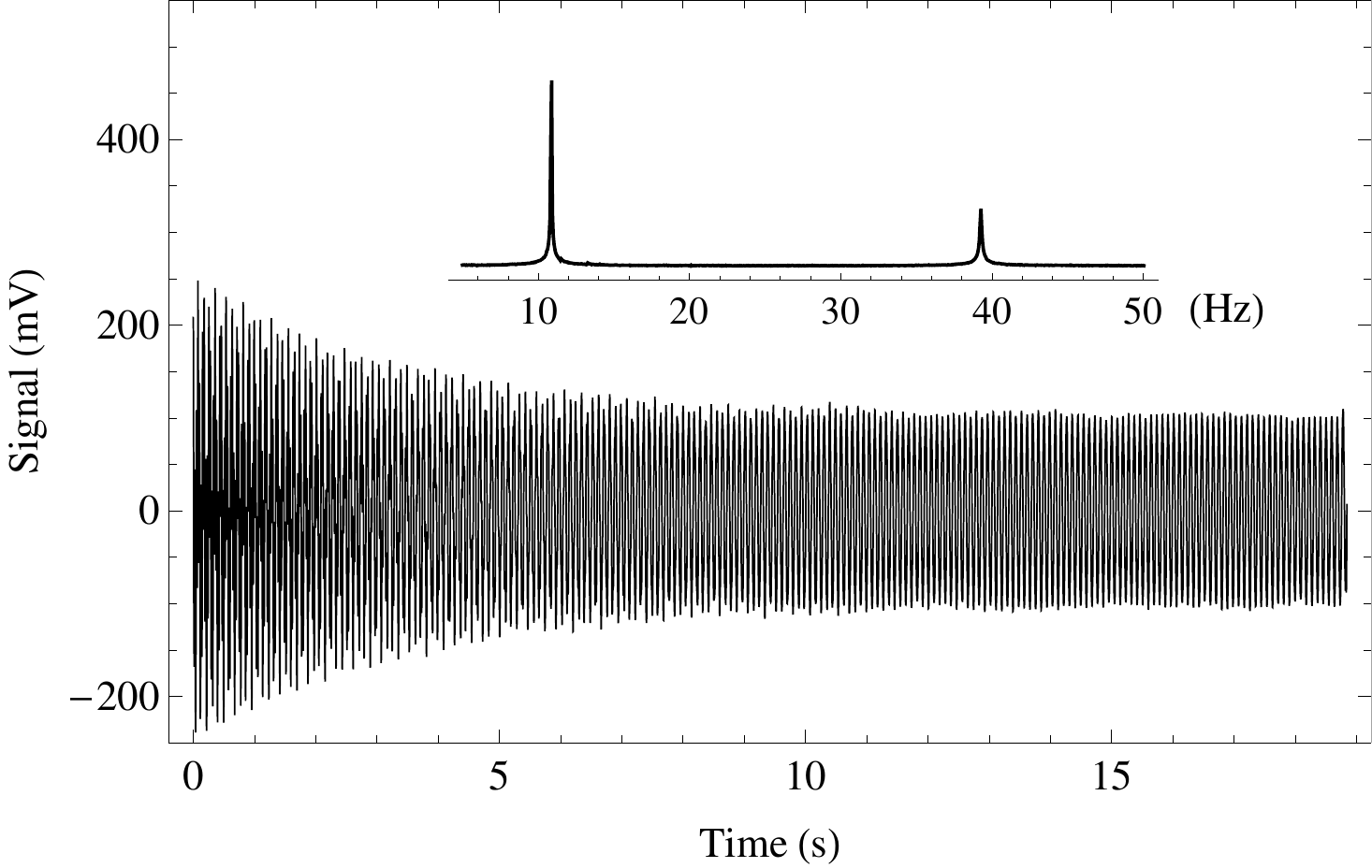}\\
  \caption{Ultra-low-field NMR signals obtained in a mixture of hyperpolarized liquid xenon and pentane.  These data were acquired in a 10 mG magnetic field, using superconducting quantum interference devices rather than atomic magnetometers.  The inset shows the magnitude Fourier transform. }\label{Fig:XePentane}
\end{figure}

We briefly present an additional comagnetometer scheme based on a mixture of hyperpolarized ${\rm ^{129}Xe}$ and pentane. Xenon is attractive for such an application because it has a very long transverse relaxation time and hyperpolarization boosts signal considerably. In the pentane-xenon mixture, protons are polarized via the spin-polarization induced nuclear Overhauser effect \cite{Navon1996,Heckman2003}, eliminating the need for a prepolarizing magnet. Furthermore, xenon $T_1$ is so long ($\approx 400~{\rm s}$, for the conditions of measurements presented here) that many transients can be acquired in a single batch of hyperpolarized xenon.  An ultra-low-field NMR signal of a mixture of hyperpolarized liquid ${\rm ^{129}Xe}$ and pentane is presented in Fig. \ref{Fig:XePentane}, following a resonant $\pi/2$ pulse that tipped the protons into the transverse plane, and produced a small transverse excitation of the ${\rm ^{129}Xe}$ spins (1-2$^\circ$).  Spin precession was monitored by superconducting quantum interference devices in a 10 mG, magnetically shielded environment. The inset shows the magnitude Fourier transform, with the proton signal at 39 Hz and the ${\rm ^{129}Xe}$ signal at 10.5 Hz.  Proton $T_2$ was about 3.2 s for these data.  $T_2$ for xenon was too long to be measured by this data set. More details of the apparatus used for these measurements can be found in Refs. \cite{Heckman2003,Ledbetter2011b}.

We now discuss several possible applications for such liquid state comagnetometers:

\textit{Precision measurements of spin-gravity} -- There has been both experimental \cite{Venema1992,Wineland1972,Wineland1991,Heckel2008,Kimball2009} and theoretical \cite{Flambaum2009,Leitner1963} interest in the question of whether spins can couple to gravity. Such an interaction violates invariance under time-reversal (\textit{T}), equivalent to \textit{CP}, the combined symmetries of charge conjugation ($C$) and spatial inversion ($P$).  The exchange of hypothetical pseudoscalar particles \cite{Moody1984}, such as axions, with an unpolarized massive body (e.g., the earth) would lead to similar effects.  The ${\rm ^1H}$-${\rm ^{19}F}$ comagnetometer outlined here could be employed for such tests by configuring the apparatus so that the Earth's gravitational acceleration $\mathbf{g}$ has a nonzero projection along the magnetic field.  In the presence of a spin-gravity coupling, the ratio of precession frequencies would be different depending on the orientation of the magnetic field.   The presently considered comagnetometer has an advantage over the ${\rm ^{199}Hg}$-${\rm ^{201}Hg}$ comagnetometer of Ref. \cite{Venema1992} in that both protons and ${\rm ^{19}F}$ nuclei are spin-1/2 particles.  In the case of ${\rm ^{201}Hg}$, with nuclear spin $I=3/2$, interaction of the nuclear quadrupole moment with the cell walls can result in systematic effects.  Reaching sensitivity of ${\rm 10^{-11}~Hz}$ would represent about four orders of magnitude improvement over the limits of Ref. \cite{Venema1992}. Since the ${\rm ^{19}F}$ nucleus has an unpaired neutron, the experiment would be sensitive to a spin-gravity coupling with a linear combination of proton and neutron spin.

\textit{Permanent electric dipole moments} -- A permanent electric dipole moment of an atom or an elementary particle also violates invariance under $T$ and $CP$, and has long been hailed as an unambiguous signature of new physics beyond the standard model.  Present experimental limits on the electric dipole moment of the electron \cite{Regan2002}, neutron \cite{Baker2006}, and ${\rm ^{199}Hg}$ atom \cite{Griffith2009} have placed constraints on many proposed extensions to the standard model \cite{Pospelov2005}.  The ${\rm ^1H}$-${\rm ^{129}Xe}$ comagnetometer scheme outlined here could be used for such an experiment by applying an electric field either parallel or antiparallel to the magnetic field.  An electric dipole moment interacts with an electric field via $H_{\rm EDM} = d\frac{\mathbf{\hat{s}}}{s}\cdot \mathbf{E}$.  Reversal of $\mathbf{E}$ gives rise to a frequency shift $\delta\nu_{Xe} = 4dE/h$.  \textit{CP} violating effects are strongly enhanced in heavy nuclei, so the proton-precession frequency would be used to compensate for magnetic field fluctuations.  In addition to previously mentioned attributes, xenon is appealing for this application because it has a very high electric field breakdown strength, on the order of 400 kV/cm \cite{Derenzo1974}, and the hydrocarbons we discuss here also have high electric field breakdown strength.  With $E=400~{\rm kV/cm}$, frequency resolution of $4\times 10^{-11}~{\rm Hz}$ corresponds to an EDM limit of approximately ${\rm 1.3\times 10^{-31}~e\cdot cm}$.  This is roughly two orders of magnitude better than the ${\rm 1.3\times 10^{-29}~e\cdot cm}$ statistical sensitivity of the ${\rm ^{199}Hg}$ EDM experiment of Ref. \cite{Griffith2009}, and four orders of magnitude better than the limit set by a gas phase ${\rm ^3He}$-${\rm ^{129}Xe}$ comagnetometer \cite{Rosenberry2001}, each obtained over months of integration.  Liquid ${\rm ^{129}Xe}$ (neat) was considered for an electric dipole moment experiment in the past \cite{Romalis2001}.  Such experiments have been complicated by nonlinear effects due to long-range dipolar fields when the spins are tipped into the transverse direction by $90^\circ$. These nonlinear effects are highly suppressed if the spins are tipped by only a few degrees \cite{Ledbetter2002}, however this precludes operation in the gradiometer mode suggested in Ref. \cite{Romalis2001}.  The presence of a second nuclear spin species would allow one to operate in the small tip angle regime, while retaining a second channel to compensate for magnetic field fluctuations.

\textit{Gyroscopes} -- Since the spins define an inertial reference frame, they can be used to sense rotations.  Sensitivity to rotations at the level of $5\times 10^{-9}~{\rm Hz}$ in single 10 second measurement would form a gyroscope competitive with other technologies based on cold atoms, ring lasers, and overlapping ensembles of electron and nuclear spins \cite{GyroRefs}

In conclusion, we have demonstrated operation of liquid-state nuclear-spin comagnetometers based on mixtures of mutually miscible solvents, each rich in a different spin species.  We have outlined how such a  device could be used for precision measurements such as a test of spin-gravity coupling and a search for permanent electric dipole moments.  Estimates based on signal-to-noise ratio for realistic conditions indicate that such devices may be two to four orders of magnitude more sensitive than previous experiments.

This research was supported by the National Science Foundation under award \#CHE-0957655 (D. Budker and M. P. Ledbetter), by the U.S. Department of Energy, Office of Basic Energy Sciences, Division of Materials Sciences and Engineering under Contract No. DE-AC02-05CH11231 (J.W. Blanchard and A. Pines), and by the Kolumb program of the Foundation for Polish Science (S. Pustelny).  We thank S. Knappe and J. Kitching for supplying the microfabricated alkali vapor cell.


\begin{thebibliography}{99}

\bibitem{Venema1992} B. J. Venema, P. K. Majumder, S. K. Lamoreaux, B. R. Heckel, and E. N. Fortson, Phys. Rev. Lett. \textbf{68}, 135-138 (1992).

\bibitem{Bear2000} D. Bear, R. E. Stoner, R. L. Walsworth, V. A. Kosteleck\'y, Phys. Rev. Lett. \textbf {85}, 5038 (2000).

\bibitem{Rosenberry2001} M. E. Rosenberry and T. E. Chupp, Phys. Rev. Lett. \textbf{86}, 22-25 (2001).

\bibitem{Regan2002} B. C. Regan, E. D. Commins, C. J. Schmidt, D. DeMille, Phys. Rev. Lett. \textbf{88}, 071805 (2002).

\bibitem{Griffith2009} W. C. Griffith \textit{et al.} Phys. Rev. Lett. \textbf{102}, 101601 (2009).

\bibitem{Gemmel2010} C. Gemmel \textit{et al.}, Euro. Phys. J. D \textbf{57}, 303 (2010).

\bibitem{Kornack2005} T. W. Kornack, R. K. Ghosh, and M. V. Romalis, Phys. Rev. Lett. \textbf{95}, 230801 (2005).

\bibitem{Packard1954} M. Packard and R. Varian, Phys. Rev. \textbf{93}, 941-941 (1954).

\bibitem{Waters1958} G. S. Waters and P.D. Francis, J. Sci. Inst. \textbf{35}, 88-93 (1958).

\bibitem{Stuart1972} W. F. Stuart, Rep. Prog. Phys. \textbf{35}, 803-881 (1972).

\bibitem{Ledbetter2011} M. P. Ledbetter \textit{et al.}, Phys. Rev. Lett. \textbf{107}, 107601 (2011).

\bibitem{Happer1973} W. Happer and H. Tang, Phys. Rev. Lett. \textbf{31}, 273 (1973).

\bibitem{Kominis2003} I. K. Kominis, T. W. Kornack, J. Allred, and M. V. Romalis, Nature \textbf{422}, 596 (2003).

\bibitem{Kornack2007} T. W. Kornack \textit{et al}., Appl. Phys. Lett. \textbf{90}, 3 (2007).

\bibitem{Yashchuk2004} V. V. Yashchuk \textit{et al.}, Phys. Rev. Lett. \textbf{93}, 160801/1 (2004).

\bibitem{Patrick1960} C. R. Patrick and G. S. Prosser, Nature \textbf{187}, 1021 (1960).

\bibitem{CRC} \textit{Handbook of chemistry and physics, 66th ed.} Eds. R. C. Weast, M. J. Astle, W. H. Beyer, (CRC press, Boca Raton, Florida).

\bibitem{Romalis2001} M. V. Romalis and M. P. Ledbetter, Phys. Rev. Lett. \textbf{87}, 067601 (2001).

\bibitem{Navon1996} G. Navon, Y.-Q. Song, T. Room, S. Appelt, R. E. Taylor,
and A. Pines, Science \textbf{271}, 1848 (1996).

\bibitem{Heckman2003} J. J. Heckman, M. P. Ledbetter, and M. V. Romalis, Phys. Rev. Lett. \textbf{91}, 067601 (2003).

\bibitem{Ledbetter2011b} M. P. Ledbetter, G. Saielli, N. Tran, A. Bagno, and M. V. Romalis, arXiv:1112.5644.

\bibitem{Heckel2008} B. R. Heckel \textit{et al.} Phys. Rev. D \textbf{78}, 092006 (2008).

\bibitem{Wineland1972} D. J. Wineland and N. F. Ramsey, Phys. Rev. A \textbf{5}, 821
(1972).

\bibitem{Wineland1991} D. J. Wineland \textit{et al.}, Phys. Rev. Lett. \textbf{67}, 1735 (1991).

\bibitem{Kimball2009} D. F. Jackson Kimball, L. R. Jacome, S. Guttikonda, E. J. Bahr, and
L. F. Chan, J. Appl. Phys \textbf{109} 063113 (2009).

\bibitem{Leitner1963} J. Leitner and S. Okubo, Phys. Rev. \textbf{136}, B1542 (1964); N.
D. Hari Dass, Phys. Rev. Lett. \textbf{36}, 393 (1976); W. T. Ni, Phys. Rev. Lett. \textbf{38}, 301 (1977).

\bibitem{Flambaum2009} V. Flambaum, S. Lambert, and M. Pospelov, Phys. Rev. D \textbf{80}, 105201 (2009).

\bibitem{Baker2006} C. A. Baker \textit{et al.}, Phys. Rev. Lett. \textbf{97}, 131801 (2006).

\bibitem{Moody1984} J. E. Moody and F. Wilczek, Phys. Rev. D \textbf{30}, 130 (1984).

\bibitem{Pospelov2005} M. Pospelov and A. Ritz, Ann. Phys. \textbf{318}, 119 (2005).

\bibitem{Derenzo1974} S. E. Derenzo, T. S. Mast, H. Zaklad, and R. A. Muller, Phys. Rev. A \textbf{9}, 2582 (1974).

\bibitem{Ledbetter2002} M. P. Ledbetter and M. V. Romalis, Phys. Rev. Lett. \textbf{89}, 287601 (2002).


\bibitem{GyroRefs} C. Jentsch, T. M\"uller, E.M. Rasel, and W. Ertmer, Gen.
Relativ. Gravit. \textbf{36}, 2197 (2004); S. J. Sanders, L. K. Strandjord, and D. Mead, in 15th
Optical Fiber Sensors Conference Tech. Digest (IEEE, Portland, 2002), p. 5; G. E. Stedman, K. U. Schreiber, and H. R. Bilger, Classical Quantum Gravity \textbf{20}, 2527 (2003); T. L. Gustavson, A. Landragin, and M. A. Kasevich, Classical Quantum Gravity \textbf{17}, 2385 (2000); T. W. Kornack, R. K. Ghosh, and M. V. Romalis, Phys. Rev. Lett. \textbf{95}, 230801 (2005).


\end{thebibliography}
\end{document}